\documentclass[conference]{IEEEtran}
\IEEEoverridecommandlockouts
\usepackage{geometry}     
\usepackage{cite}
\usepackage{amsmath,amssymb,amsfonts}
\usepackage{booktabs}
\usepackage{algorithmic}
\usepackage{graphicx}
\usepackage{textcomp}
\usepackage{xcolor}
\usepackage{hyperref}
\usepackage{makecell}
\usepackage[linesnumbered,ruled,vlined]{algorithm2e}
\SetKwInput{KwInput}{Input}                
\SetKwInput{KwOutput}{Output}              
\begin{document}

\title{LlamaF: An Efficient Llama2 Architecture Accelerator on Embedded FPGAs\\
}

\author{
    \begin{tabular}{cccc}
        \begin{minipage}{0.23\textwidth}
            \centering
            \footnotesize
            Han Xu \\
            \textit{University of Illinois at Urbana-Champaign} \\
            Urbana, Illinois \\
            hanxu8@illinois.edu
        \end{minipage}
        &
        \begin{minipage}{0.23\textwidth}
            \centering
            \footnotesize
            Yutong Li \\
            \textit{University of Illinois at Urbana-Champaign} \\
            Champaign, Illinois \\
            yutong23@illinois.edu
        \end{minipage}
        &
        \begin{minipage}{0.23\textwidth}
            \centering
            \footnotesize
            Shihao Ji \\
            \textit{Georgia State University} \\
            Atlanta, Georgia \\
            sji@gsu.edu
        \end{minipage}
    \end{tabular}
}


\maketitle
\author{\IEEEauthorblockN{Anonymous Authors}} 

\begin{abstract}
Large language models (LLMs) have demonstrated remarkable abilities in natural language processing. However, their deployment on resource-constrained embedded devices remains difficult due to memory and computational demands. In this paper, we present an FPGA-based accelerator designed to improve LLM inference performance on embedded FPGAs. We employ post-training quantization to reduce model size and optimize for off-chip memory bandwidth. Our design features asynchronous computation and a fully pipelined accelerator for matrix-vector multiplication. Experiments of the TinyLlama 1.1B model on a Xilinx ZCU102 platform show a 14.3-15.8x speedup and a 6.1x power efficiency improvement over running exclusively on ZCU102 processing system (PS).
\end{abstract}


\section{Introduction}
Recent advancements in large language models (LLMs) have significantly impacted natural language processing (NLP) and a range of interdisciplinary areas. These models, capable of generating human-like interactions, are poised to transform a myriad of industries, including healthcare~\cite{LLMHealth} biomedical~\cite{liu2024timemattersexaminetemporal}, entertainment~\cite{ding-24-llava}, robotics~\cite{zeng2023largelanguagemodelsrobotics} and music~\cite{deng2024composerx}. However, deploying these models on IoT and embedded systems presents a significant challenge due to their extensive matrix operations that surpass the computational and memory capacities of these device~\cite{10285619}.

Field Programmable Gate Arrays (FPGAs), known for their reconfigurable architecture and energy efficiency, offer a promising solution to accelerate LLMs on IoT devices. Despite their potential, the substantial memory requirements of models like the 4.4GB TinyLlama 1.1B~\cite{zhang2024tinyllama} often surpass the memory capacity of FPGA boards, such as the Xilinx ZCU102 with 4GB off-chip and 5.11MB on-chip memory.

To address these challenges, quantization has been leveraged to compress model weights while incurring insignificant losses of model's predictive performances. Considering that quantized models are stored in off-chip DDR and transferred to FPGAs via off-chip memory interfaces, three key challenges arise in accelerating the inference of LLMs on FPGAs: (1) effectively utilizing the off-chip memory bandwidth, (2) a fully pipelined FPGA accelerator to optimize data throughput, and (3) minimizing memory overhead while loading billions of model weights. In this paper, we address these challenges with LlamaF, an efficient Llama2~\cite{llama2} architecture accelerator designed for embedded FPGAs. To the best of our knowledge, this is the first work that
accelerates the Llama2 architecture on embedded FPGAs. Our key contributions are summarized below:
\begin{itemize}
    \item We substantially reduce off-chip DDR memory bandwidth requirements through post-training quantization of LLM weights and run-time quantization of inference parameters.
    \item We propose a fully pipelined accelerator for group-wise quantized matrix-vector multiplication (GQMV).    
    \item We enable asynchronous FPGA computation during weight transfer within each layer, which leads to notable performance gains.
    \item We demonstrate the effectiveness of LlamaF by accelerating the TinyLlama 1.1B model on the Xilinx ZCU102 platform. Experimental results reveal a 14.3-15.8x speedup and a 6.1x power efficiency improvement compared to running exclusively on the ZCU102 processing system (PS).
\end{itemize}

\section{Background and related work}

\subsection{Transformer and Llama2}
The transformer~\cite{vaswani2023attention} is a deep learning architecture that utilizes a self-attention mechanism to map relationships between inputs and outputs, offering significant advantages in NLP tasks, including sentiment analysis~\cite{mo2024finetuninggemma7benhancedsentiment}, question answering~\cite{jiang2024multi}, social computing~\cite{influence-pathway-social}, information extraction~\cite{li-etal-2023-open} and text generation~\cite{liu-etal-2023-ask}. It is also the de-facto architecture of many state-of-the-art LLMs, which are capable of generating text from encoded prompts consisting of a sequence of tokens.

While batch processing (i.e., handling multiple tokens simultaneously) enhances computational efficiency, it increases the latency, which is critical for real-time processing on embedded devices. Hence, this paper adopts a batch size of one to meet the real-time processing requirements.

During text generation, a forward pass computes logits for each token to facilitate contextual understanding. These logits represent scores associated with the next potential token. If a token falls within the user's prompt range, the model follows the prompt; otherwise, it uses a sampling strategy such as top-p~\cite{Holtzman2020Top-p} or greedy sampling to select the next token. Text generation terminates if the model selects an end-of-sentence (EOS) token. If not, the token is mapped back to text and is printed out.

\begingroup
\setlength{\belowcaptionskip}{-5pt}
\begin{figure}[h]
  \centering
  \includegraphics[width=0.95\linewidth]{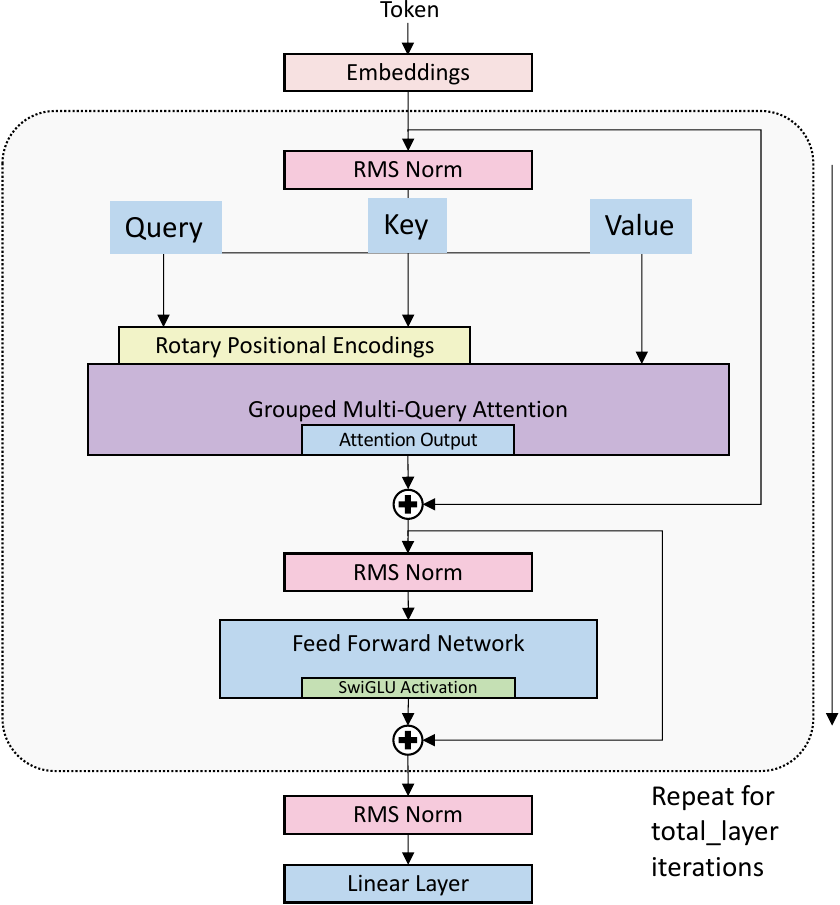}\vspace{-5pt}
  \caption{Llama2 Architecture: Forward Pass.}\label{fig:llama2}
\end{figure}
\endgroup

Llama2 extends the transformer architecture with innovations such as grouped-query attention (GQA)~\cite{ainslie-etal-2023-gqa} and Rotary Position Embedding (RoPE)~\cite{rope}, which enhance handling of multi-head attention and long sequences, respectively. As depicted in Fig.~\ref{fig:llama2}, the forward pass of Llama2 processes the input tokens by referencing its associated embedding. This data flows sequentially through multiple layers (22 in TinyLlama). Each layer includes root mean square normalization (RMSNorm)~\cite{zhang-sennrich-rmsnorm}, QKV (Query, Key, Value) projections, attention calculations enhanced by RoPE and GQA, and a feed-forward network (FFN) using the SwiGLU activation function~\cite{shazeer2020swiglu}. Following this, a final linear layer outputs the logits.

Consider a single token as input, which is represented as a $dim$-dimensional embedding vector $x \in \mathbb{R}^{dim \times 1}$, e.g., $dim = 2048$ for TinyLlama. The dimensions of the Llama2 weight matrices are provided in Table~\ref{table:llama-weight-matrix}. Taking $W_q \in \mathbb{R}^{dim \times dim}$ and vector $x \in \mathbb{R}^{dim \times 1}$ as an example, we can compute the output vector $out \in \mathbb{R}^{dim \times \text{1}}$ using matrix-vector multiplication: $out = W_qx$.

\begin{table}[!htb]
\centering
\caption{Llama2 Weight Matrix Specifications}\label{table:llama-weight-matrix}
\begin{tabular}{|c|c|c|c|}
\hline
\textbf{Matrix} & \textbf{Dimension}              & \textbf{Type}  & \makecell{\textbf{Quan-}\\\textbf{tized}} \\ \hline
$W_{embeddings}$           & $(vocab\_size, dim)$       & Embeddings & Yes \\ \hline
$W_{classifier}$           & $(vocab\_size, dim)$      & Classifier & Yes \\ \hline

$W_q, W_o$           & $(dim, dim)$       & Self-Attention & Yes \\ \hline

$W_k, W_v$           & $(kv\_dim, dim)$ & Self-Attention & Yes \\ \hline

$W_1, W_3$           & $(hidden\_dim, dim)$ & FFN  & Yes \\ \hline

$W_2$           & $(dim, hidden\_dim)$ & FFN  & Yes \\ \hline

$W_{att\_norm}$           & $(dim, 1)$      & RMSNorm & No \\ \hline

$W_{ffn\_norm}$           & $(dim, 1)$      & RMSNorm & No \\ \hline

$W_{norm\_output}$           & $(dim, 1)$      & RMSNorm & No \\ \hline
\end{tabular}
\end{table}

\subsection{Quantization}
Quantization is a process that maps a broad set of values to a more compact set, which is essential for efficient numerical representation during neural network training and inference under resource constraints. A common approach in modern machine learning is the symmetric INT8 quantization, defined by
\begin{equation}
\label{eq:quant}
    Q(r) = Int(r/S),
\end{equation}
where $Q$ represents the quantization operator, $r$ is an input of real value, and $S$ is set to $\frac{2max(|r|)}{255.0}$ to utilize the full INT8 range of [-128, 127]. Furthermore, the $Int()$ function rounds or truncates a value to its nearest integer. The dequantized value, denoted as $\hat{r}$, can be recovered using
\begin{equation}
\label{eq:dequant}
    \hat{r} = Q(r) \times S.
\end{equation}
This approach, however, is prone to distortion from outliers, which extend the range of quantized values and reduce the quantization precision\cite{gholami2021survey}.

To mitigate this issue, the layer-wise quantization utilizes a uniform range across layers, but still falls short in capturing the nuances in weight distribution. To further improve the accuracy, the group-wise quantization~\cite{DBLP:conf/aaai/ShenDYMYGMK20} divides each weight matrix into groups of a predefined group size ($GS$), assigning each group a unique quantization range. This fine-grained approach enhances the accuracy of weight representations without increasing the hardware complexity significantly.

\begin{algorithm}
\caption{Group-wise Quantized Matrix-Vector Multiplication (GQMV)}\label{algo:quantization}
\KwInput{Quantized matrix $wq$ and its scaling factors $ws$, quantized vector $xq$ and its scaling factors $xs$, matrix row size $m$, matrix column size $n$, group size $GS$}
\KwOutput{Output array $out$}
\BlankLine
Initialize $out$, $ws_{cnt} = 0$\;
\For{$i \leftarrow 0$ \textbf{to} $m-1$}{
    Initialize $sum = 0$, $xs_{cnt} = 0$\;
    $offset = i \times n$\;
    \For{$j \leftarrow 0$ \textbf{to} $n-1$ \textbf{in steps of} $GS$}{
        $group\_sum = 0$\;
        \For{$k \leftarrow 0$ \textbf{to} $GS-1$}{
            $group\_sum \mathrel{+}= xq[j + k] \times wq[offset + j + k]$\;
        }
        $sum \mathrel{+}= group\_sum \times ws[ws_{cnt}++] \times xs[xs_{cnt}++]$\;
    }
    $out[i] \leftarrow sum$\;
}
\end{algorithm}

Algorithm~\ref{algo:quantization} performs matrix-vector multiplication using an 8-bit weight and 8-bit activation (W8A8) quantization scheme. The weight matrix $W$ and input vector $x$ are quantized into flatten arrays $wq$ and $xq$ respectively, with scaling factors stored sequentially in arrays $ws$ for weights and $xs$ for the input vector. The computation iterates over the weight matrix row-by-row (within the $i$ loop). For each row, it processes elements in groups of size $GS$ (in the $j$ loop). For each group, the products of corresponding elements from $wq$ and $xq$ are accumulated as a $group\_sum$ (in the $k$ loop). This $group\_sum$ is then adjusted by scaling factors from $ws$ and $xs$, and aggregated to derive the final $sum$ for each row, stored in the output array ($out$).

\subsection{Profiling TinyLlama}\label{sec:profile}

We profile the TinyLlama model running solely on ZCU102 PS, using OpenMP to utilize all four A53 ARM cores. Our analysis focuses on the model's forward pass, specifically examining the runtime distribution among various computation components (in Fig.~\ref{fig:llama2}) during the generation of the 64th, 128th, 256th tokens (indexed as 63, 127, and 255, respectively due to zero-based indexing).

\begin{table}[!htb]
\centering
\caption{Llama2 Forward Pass Profiling (TinyLlama model)}\label{table:profile}
\begin{tabular}{|c|c|c|c|}
\hline
\textbf{Computation} & \textbf{pos=63} & \textbf{pos=127} & \textbf{pos=255} \\ \hline
Matrix Computation & \textbf{98.98\%} & \textbf{98.53\%} & \textbf{97.64\%} \\ \hline
Multi-head Attention & 0.47\% & 0.92\% & 1.82\% \\ \hline
SwiGLU & 0.13\% & 0.13\% & 0.13\% \\ \hline
RoPE & 0.07\% & 0.07\% & 0.07\% \\ \hline
RMSNorm & 0.06\% & 0.06\% & 0.05\% \\ \hline
\end{tabular}
\end{table}

Table~\ref{table:profile} reveals that while the runtime for multi-head attention slightly increases with the position, matrix computations overwhelmingly dominate the runtime. In all scenarios, these computations consistently account for more than 97\% of the total runtime. This underscores the importance of accelerating matrix multiplications on FPGAs, as indicated by the blue modules in Fig.~\ref{fig:llama2}.

\subsection{Related Work}
Utilizing FPGAs to accelerate LLMs has gained significant attention due to the energy efficiency and suitability of FPGAs for edge computing. An  example is FQ-BERT~\cite{liu2021hardware}, which demonstrates the feasibility of running the BERT model~\cite{devlin-etal-2019-bert} on FPGAs using a W4A8 quantization scheme. However, BERT utilizes an encoder-only transformer architecture, while the dominant trend in LLMs is towards decoder-only models. These models offer significant advantages in scalability and impressive zero-shot performance and few-shot learning capabilities~\cite{brown2020language}.

More recently, Chen et al. explore model-specific spatial acceleration for LLM inference on FPGAs~\cite{chen2023understanding}. Their approach specializes hardware units for specific operators or layers, which enables direct communication within a dataflow architecture. However, their implementation is limited to BERT base (110M parameters) and GPT2 Medium (355M parameters)~\cite{radford2019-gpt2}. These models are significantly smaller than current LLMs, which often exceed billions of parameters.

FlightLLM~\cite{zeng2024flightllm} facilitates efficient LLM inference on FPGAs using a configurable sparse DSP chain and an always-on-chip decode scheme. Their implementation runs OPT-6.7B~\cite{zhang2022optopenpretrainedtransformer} and LlaMA2-7B on Xilinx Alveo U280 FPGA. However, the U280 FPGA is designed for data centers. The field of accelerating LLM inference specifically on embedded FPGAs remains largely unexplored.

\section{Quantization and Software Design}
In this section, we first present the quantization strategy used in the paper. Then, we demonstrate the software design choices applied on the host side (PS subsystem) of embedded FPGA.

\subsection{Quantization}\label{AA}
To ensure the accuracy of the quantized TinyLlama model, we employ a group size of 256 ($GS\!=\!256$). This size is selected based on its compatibility with the dimensional parameters of TinyLlama, all of which are divisible by 256. We implement a W8A8 quantization scheme (Eq.~\ref{eq:quant}) to the model's original 32-bit floating-point weight/activation format (W32A32). The quantization process primarily targets the weights of the token embedding, attention, FFN, and classifier layers due to their substantial impact on model size. Normalization weights are excluded from quantization, given their smaller size leading to negligible overhead (Table~\ref{table:llama-weight-matrix}). This strategy effectively reduces the model size from 4.4GB to 1.1GB.

\begin{figure*}[htbp]
  \centering
  \includegraphics[width=\textwidth]{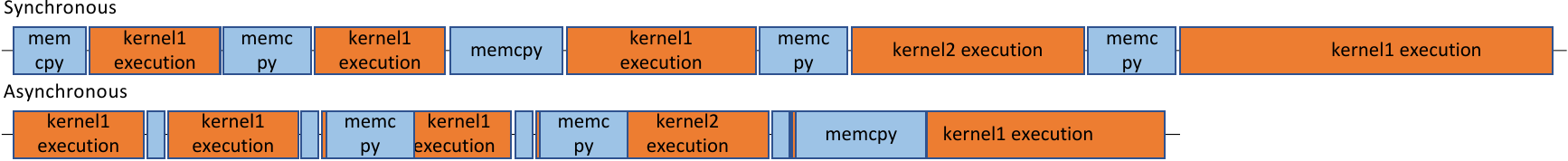}
  \caption{Comparison of synchronous vs. asynchronous FPGA computation.}\label{fig:schedule}
\end{figure*}

\subsection{Software Design}\label{sec:schedule}
Algorithm~\ref{alg:host} presents the forward pass of the FPGA-accelerated Llama2 architecture, expanding upon the concept illustrated in Fig.~\ref{fig:llama2}. The transformer controller with KV caches runs on the PS, with the quantized weights and activations transmitted from the PS to the FPGA Programmable Logic (PL). To minimize the kernel launch overhead, we concatenate weight matrices that share the same input vector. This includes concatenating $W_q$, $W_k$, and $W_v $ (line 4), as well as $W_1$ and $W_3$ (line 12). The output vector is then decoupled following the matrix-vector multiplication.

We retain the computation of multi-head attention on the PS due to the complexities of accelerating softmax operations on FPGAs. Furthermore, Llama2 profiling from ~\ref{sec:profile} suggests that multi-head attention contributes minimal overhead with smaller step sizes. To optimize performance, we employ OpenMP to parallelize the computation.

The Llama2 architecture utilizes two distinct matrix column sizes in its matrix-vector multiplications - $dim$ and $hidden\_dim$. For most computations, the column size aligns with the input dimension $dim$ (2,048 for TinyLlama, and 8 groups due to $GS\!=\!256$). However, the final FFN matrix-vector computation projects it back to the original input dimension, requiring the column size matches the $hidden\_dim$ (5,632 for TinyLlama, with 22 groups due to $GS=256$).

To accommodate the varying group sizes, we have implemented separate GQMV FPGA kernels ($kernel1$ and $kernel2$). To minimize DDR memory consumption, weights for each layer are loaded into the buffer sequentially rather than simultaneously. This method requires only 111.5 MB of buffer space, as opposed to the 1.1 GB that would be needed if all layers were loaded at once. Furthermore, we enhance the efficiency of the forward pass through task-level scheduling, allowing off-chip parameter transfers to overlap with kernel execution (Fig.~\ref{fig:schedule}). Specifically, 
\begin{itemize}
    \item First-layer ($l = 0$) weights: For the very first forward pass, the buffers are initialized and loaded at program start.
    \item Subsequent ($l > 0$) weights: After finishing a GQMV operation, the next layer's corresponding weights are loaded asynchronously during the kernel execution.
\end{itemize}

This strategy effectively masks memory transfer time within the kernel execution, significantly reducing the overhead associated with off-chip transfers.

\begin{algorithm}
\caption{FPGA-Accelerated Transformer Forward Pass Host-Code}\label{alg:host}
\KwInput{Input Token $token$, Position $pos$, FPGA Kernels $kernel1$ and $kernel2$, $W_{embeddings},$ $W_q, W_k, W_v, W_1, W_2, W_3, W_{classifier}$,
TinyLlama Vocabulary Size $vocab\_size$,
Input Dimension $dim$, FFN's Hidden Dimension $hidden\_dim$, Key/Query Dimension $kv\_dim$
}
\KwOutput{Logits vector $logits$}
Copy token embedding to input vector $x$ based on $token$\;
\For{each layer $l$ in $total\_layers$}{
RMSNorm and quantize $x$\;
$q,k,v \leftarrow$ kernel1($x$, $W_q$ + $W_k$ + $W_v$, $dim + kv\_dim + kv\_dim$)\;
Apply RoPE positional encoding to $q$ and $k$ using position $pos$\;
// Perform multi-head attention in parallel\;
$att\_buffer \leftarrow \text{multi-head\_att}(q, k, v, \text{pos})$\;
Quantize $att\_buffer$\;
$att\_output \leftarrow$ kernel1($att\_buffer$, $W_o$, $dim$)\;
Residual connection from $att\_output$\;
RMSNorm and quantize $x$\;
$ffn\_buffer \leftarrow$ kernel1($x$, $W_1$ + $W_3$, $hidden\_dim + hidden\_dim$)\;
SwiGLU and quantize $ffn\_buffer$\;
$ffn\_output \leftarrow$ kernel2($ffn\_buffer$, $W_2$, $dim$)\;
Residual connection from $ffn\_output$\;
}
RMSNorm and quantize $x$\;
$logits \leftarrow$ kernel1($x$, $W_{classifier}$, $vocab\_size$)\;
\BlankLine
Return $logits$\;
\end{algorithm}

\section{Hardware Design}\label{sec:hardwaredesign}
\subsection{Overview}

\begingroup
\setlength{\belowcaptionskip}{-5pt}
\begin{figure*}[h]
  \centering
  \includegraphics[width=0.98\linewidth]{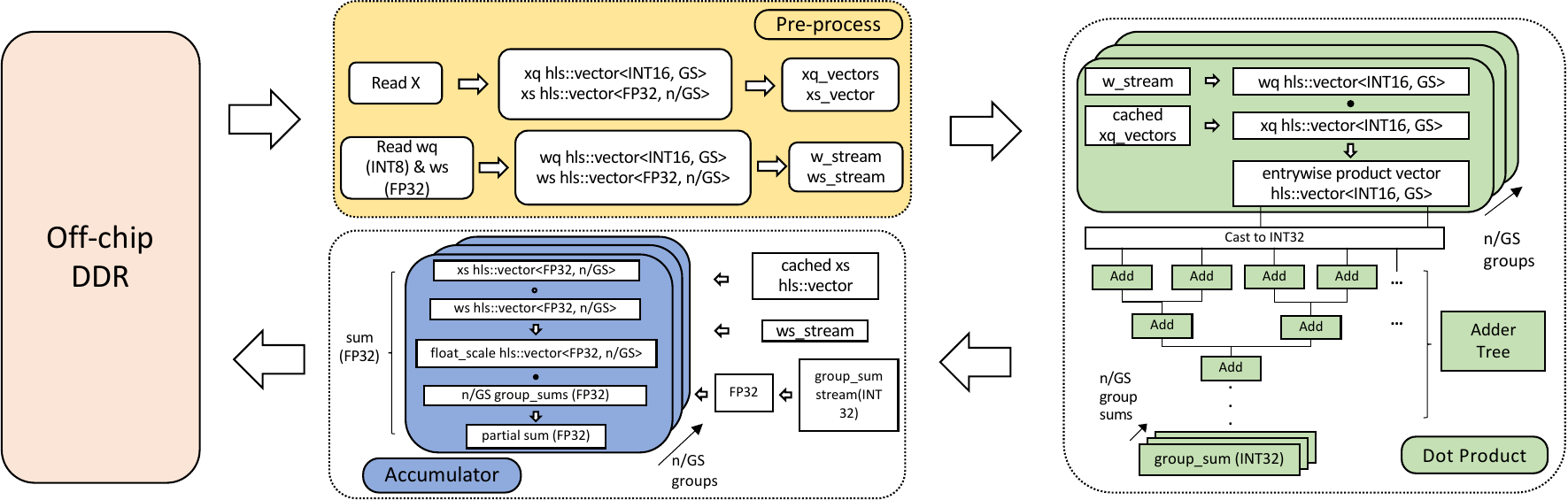}
  \caption{Overview of LlamaF hardware design.}\label{fig:accelerator}
\end{figure*}
\endgroup

As shown in Fig.~\ref{fig:accelerator}, the hardware accelerator is designed to perform efficient GQMV. It is structured around three principal stages: pre-processing, dot-product, and accumulate. The pre-processing stage prepares quantized weights and vectors, along with their scaling floats, for computation. The dot-product stage performs the vector-vector dot product calculation. The accumulate stage scales and aggregates the partial results, and transfers the output to the off-chip memory. After pre-fetching $x$, the accelerator uses $dataflow$ to facilitate task-level pipelining among stages, thereby enhancing concurrency and optimizing workload distribution across data reading, computation, and output writing.

\subsection{Pre-processing Stage}
The pre-processing stage begins by reading the quantized vector $x$, casting INT8 to INT16, and storing the quantized values ($xq$) and its scales ($xs$) in \textit{hls::vector} within the FPGA's Block RAM (BRAM). This caching strategy minimizes reads in subsequent iterations. Additionally, the stage processes the quantized weight and its scale values iteratively. During each iteration, the inputs are streamed, ready to be consumed by the later stages:
\begin{itemize}
    \item Quantized weights: The process initially reads INT8 weight elements from $wq$. Similar to casting $xq$, these elements are converted to INT16 to ensure sufficient range for the dot product calculations. This step is critical as the product of two INT8 values can exceed the INT8 range, potentially causing overflow. After conversion, the weights are grouped into \textit{hls::vector} of size $GS$. These grouped weights are then directed to $w\_stream$, ensuring a continuous and efficient flow of data to the subsequent dot-product stage.
    \item Weight scale values: Simultaneously, the pre-processing stage retrieves the corresponding scale values from $ws$. The scale values are packed into \textit{hls::vector} of size $n/GS$, and forwarded via $ws\_stream$.
\end{itemize}

\subsection{Dot-product Stage}
In each iteration $i$, the dot-product stage processes $n/GS$ groups. It receives \textit{hls::vector} of $wq$ from the $w\_stream$ and fetches the corresponding \textit{hls::vector} for activation $x$ from the same quantization group. The stage performs SIMD vector multiplication on these vector pairs to create temporary multiplication vectors. Next, it applies a reduction-add operation across all elements within each temporary vector (i.e., within the same group) using an adder tree (of depth 8 due to $GS\!=\!256$). The adder tree's first layer casts INT16 to INT32 to prevent overflow. This operation yields the dot product of the two vectors as a group sum. Given that there are $n/GS$ groups in each iteration, an equal number of group sums are forwarded to $group\_sum\_stream$.

\subsection{Accumulate Stage}
During each iteration $i$, the accumulate stage fetches a \textit{hls::vector} from $ws\_stream$, and a \textit{hls::vector} from $x\_scale$. It first performs element-wise multiplication between the two vectors, each of size $n/GS$, resulting in a $float\_scale$ vector of the same size. Subsequently, the accumulate stage fetches $n/GS$ INT32 group sums from $group\_sum\_stream$. To accomodate computation with FP32 scale values, INT32s are casted to FP32. To optimize the operations within specific bit ranges and fit in the limited hardware resources, we gradually cast the intermediate values from INT8 to INT16, then from INT16 to INT32, and finally from INT32 to FP32. Afterwards, the $float\_scale$ vector undergoes a dot product operation with the corresponding FP32 group sum values. The result of the product - a scalar $sum$ - is then transmitted to the off-chip DDR memory, where it is stored at index $i$ within the output vector $out$.

\begin{algorithm}
\caption{GQMV accelerator using HLS}
\KwInput{Input quantized vector $x$, weight matrix $wq$, weight scaling factors $ws$, matrix row size $m$}
\KwOutput{Output vector $out$}
\BlankLine
Initialize $w\_stream$, $ws\_stream$\;
Initialize $group\_sum\_stream$\;
$xq\_vectors, xs\_vector \leftarrow$ pre\_fetch($x$);
\BlankLine
\#pragma HLS DATAFLOW\;
\For{$i \leftarrow 0$ \KwTo $m-1$}{
    $w\_stream \leftarrow$ read\_cast($wq$, $i$)\;
    $ws\_stream \leftarrow$ read\_scale($ws$, $i$)\;
    $group\_sum\_stream \leftarrow$ dot\_product($w\_stream$, $xq\_vectors$)\;
    $out[i] \leftarrow$ accumulate($group\_sum\_stream$, $ws\_stream$, $xs\_vector$)\;
}

\end{algorithm}
\vspace{-10pt}

\section{Experiment}

\subsection{Experiment Setup}
Our design is implemented on the Xilinx Zynq ZCU102 platform, featuring the Zynq UltraScale+ XCZU9EG MPSoC. This MPSoC includes a quad-core ARM Cortex-A53 processor with 4GB DDR4 memory on the PS, as well as PL with hardware resources detailed in Table~\ref{table:resource}. The communication between PS and PL is facilitated by high-performance (HP) AXI interfaces, each supporting a data transfer rate of up to 128 bits per second and a maximum full-duplex bandwidth of 85 Gbps.

\begin{table}[ht]
\centering
\caption{Hardware Utilization of LlamaF on ZCU102}\vspace{-5pt}
\label{table:resource}
\begin{tabular}{lcccc}
\hline
& \textbf{LUT} & \textbf{FF} & \textbf{BRAM} & \textbf{DSP} \\
\hline
ZCU102 Total & 274080 & 548160 & 912 & 2520 \\
LlamaF on ZCU102 & 59.72\% & 31.31\% & 24.45\% & 20.95\% \\
\hline
\end{tabular}
\end{table}

The host code and accelerator are both developed in C++, with the acceleator converted to Verilog using Vitis HLS 2022.2 and operating at a 205 MHz frequency. We use aarch64-xilinx-linux-g++ 11.2.0 for compiling the host binary and Vivado 2022.2 for bitstream generation. Vitis 2022.2 is used to create the platform and boot images. Table~\ref{table:resource} illustrates the resource utilization on ZCU102.\vspace{-5pt}

\subsection{Accuracy of the quantized model}

Our W8A8 quantization strategy transfers 16 8-bit values per cycle using available bandwidth and is evaluated on benchmark datasets both theoretically and practically.

\subsubsection{Quantization Error Assessment}
We analyze the impact of INT8 quantization on the TinyLlama model by calculating the quantization error for each value as follows:
\begin{equation}
    Error = |\hat{r} - r|
\end{equation}

Table~\ref{table:quant-stats} details the statistics of quantization error (in theory) across all groups. Furthermore, when considering the error percentage $\frac{Error}{r}$, the average error percentage is 3.30\%, with a standard deviation of 11.57\%.
\vspace{-10pt}
\begin{table}[htbp]
\centering
\caption{Statistics of group-wise quantization error (GS=256)}
\label{table:quant-stats}
\begin{tabular}{@{}lcccc@{}}
\toprule
Method & Max & Min & Mean  & \begin{tabular}[c]{@{}c@{}}Std \end{tabular} \\
\midrule
INT8 Quantization & 0.0115 & 0.0 & 0.000265 & 0.000173 \\
\bottomrule
\end{tabular}
\end{table}

\subsubsection{Accuracy on benchmark dataset} We further examine the impact of quantization by measuring the perplexity (PPL) on the WikiText-2 language modeling task~\cite{merity2016pointer}. According to Table~\ref{table:model-ppl}, the W8A8 quantization results in a slight PPL increase by 0.57\%, which is within the range of theoretical accuracy degradation.

\vspace{-10pt}
\begin{table}[htbp]
\centering
\caption{Comparison of TinyLlama Model PPL (lower is better)}
\label{table:model-ppl}
\begin{tabular}{@{}lcc@{}}
\toprule
Model & W32A32 PPL & W8A8 (GS=256) PPL \\
\midrule
TinyLlama & 7.05 & 7.09 \\
\bottomrule
\end{tabular}
\end{table}

\subsection{Performance Analysis}\label{sec:perf-analysis}

We evaluate our implementation against the baseline -- the same quantized TinyLlama (using the same W8A8 scheme) running exclusively on the ZCU102 PS (without FPGA acceleration). Performance is assessed by answering a subset of questions from the SQuAD dataset~\cite{rajpurkar-etal-2016-squad}. Specifically, we omit the EOS token, apply greedy sampling, and vary the step size ($64$, $128$, and $256$) to measure the tokens generated per second (tok/s). Since GQMV involves both integer and floating-point arithmetic, we average the runtime of logits computation to measure the Giga Operations Per Second (GOPS) of GQMV operations. Additionally, to evaluate energy efficiency when $step\!=\!256$, we utilize the ZCU102 system controller user interface (SCUI) to monitor power consumption.

\vspace{-10pt}
\begin{table}[htbp]
\centering
\caption{Comparison of inference speed and power consumption}
\label{table:model-performance-power}
\begin{tabular}{@{}lccccc@{}}
\toprule
Method & GOPS & \begin{tabular}[c]{@{}c@{}}step=64\\tok/s\end{tabular} & \begin{tabular}[c]{@{}c@{}}step=128\\tok/s\end{tabular} & \begin{tabular}[c]{@{}c@{}}step=256\\tok/s\end{tabular} & \begin{tabular}[c]{@{}c@{}}Efficiency\\(tok/s/W)\end{tabular} \\
\midrule
ZCU102 PS & 0.201 & 0.0935 & 0.0933 & 0.0928 & 0.0480 \\
LlamaF (no \\scheduling) & \textbf{4.696} & 0.936 & 0.915 & 0.853 & 0.207 \\
LlamaF  & \textbf{4.696} & \textbf{1.478} & \textbf{1.424} & \textbf{1.328} & \textbf{0.291} \\
  & (23.4x) & (15.8x) & (15.3x) & (14.3x) & (6.1x) \\
\bottomrule
\end{tabular}
\end{table}

The results in Table~\ref{table:model-performance-power} shows that FPGA acceleration significantly enhances inference speed and energy efficiency. Specifically, LlamaF achieves a 14.3-15.8x speedup in tok/s and a 6.1x improvement in energy efficiency compared to the ZCU102 PS. This improvement is primarily attributed to a 23.4x increase in GOPS resulting from FPGA-accelerated GQMV operations. Moreover, adding \ref{sec:schedule}'s asynchronous scheduling contributes to a 55.6-57.9\% improvement in performance over the non-scheduled version using only the hardware accelerator.

As noted in~\ref{sec:profile}, the decrease in tok/s with larger step sizes is primarily due to increased computation demands of multi-head attention. However, the acceleration of the multi-head attention on FPGAs is beyond the scope of this paper, and we leave this as a future work.

\section{Conclusion}
This paper introduces LlamaF, an efficient Llama2 architecture accelerator on embedded FPGAs. We employ a group-wise W8A8 quantization scheme to reduce memory footprint and bandwidth requirements. Additionally, we introduce a task-level scheduling to overlap off-chip parameter transfer with the FPGA kernel execution to improve the performance further. Our design features a fully pipelined accelerator that effectively balances the workload. Experimental results demonstrate a substantial 14.3-15.8x speedup in token generation and a 6.1x improvement in energy efficiency compared to running exclusively on ZCU102 PS. To the best of our knowledge, this is the first work that accelerates the Llama2 architecture inference on embedded FPGAs. As of future works, we plan to accelerate multi-head attention with a larger step size by approximating the softmax computation.

\section*{Acknowledgment}
We would like to thank the anonymous reviewers for their comments and suggestions, which helped improve the quality of this paper. We would also gratefully acknowledge the support of VMware Inc. for its university research fund to this research.

\end{document}